\documentclass[pra,aps,twocolumn,floatfix]{revtex4}
\usepackage{graphicx,graphics,psfrag,amsmath,calc}
\usepackage{epsfig}
\usepackage{color, comment}
\usepackage{mathbbol}
\usepackage{float}
\usepackage{placeins}
\usepackage{subfigure}

\begin{document}

\title{
Topological Color-Hall Insulators: \\
SU(3) Fermions in Optical Lattices}

\author{Man Hon Yau and C. A. R. S\'a de Melo}

\affiliation{
School of Physics, Georgia Institute of Technology, 
Atlanta, 30332, USA
}

\date{\today}

\begin{abstract}
We discuss the emergence of topological color insulators in optical lattices
as quantum phases of SU(3) ultra-cold neutral fermions. 
We construct the Chern matrix and classify all insulating phases in terms 
of three topological invariants: the charge-charge, the color-charge and 
the color-color Chern numbers. Our classification transcends that of 
SU(2) systems which require only the charge-charge (charge-Hall) 
and spin-charge (spin-Hall) Chern numbers. To illustrate the 
topological classification of the insulating phases of ${\rm SU(3)}$ fermions, we 
construct phase diagrams of chemical potential and color-orbit parameter 
versus color-flip fields for fixed magnetic flux ratio.
\end{abstract}

\maketitle

%
%

The identification of topological invariants for charged ${\rm SU(2)}$ fermions 
has been very important in the distinction between trivial and non-trivial
insulators found in condensed matter physics. The integer quantum Hall effect
is a typical example of the importance of topological invariants, where the 
quantum Hall conductance (charge-charge response) is proportional to an 
integer~\cite{thouless-1982} in two-dimensional 
lattices at high magnetic fields. This topological invariant, 
known as the TKKN~\cite{thouless-1982} integer, is also identified as 
the first Chern number of a ${\rm U(1)}$ 
principal fiber bundle on a torus~\cite{kohmoto-1985}. The TKKN 
integer counts the number and 
chirality of edge states in two-dimensional ${\rm SU(2)}$ systems with open 
boundary conditions, as indicated by the bulk-edge 
correspondence~\cite{hatsugai-1993, hatsugai-1996}. 

More recently, topological insulators, 
with zero quantum Hall and non-zero
quantum spin-Hall conductances, were found to exist
in graphenelike two-dimensional lattices 
without Zeeman or magnetic fields, but in the presence of spin-orbit 
coupling~\cite{kane-2005}. When spin is conserved, 
the quantization of the spin-charge 
(spin-Hall) response is the result of spin-dependent TKNN integers 
having opposite values, that is, 
${\cal C}_{\rm TKNN}^{\uparrow} = - {\cal C}_{\rm TKNN}^{\downarrow}$,
in which case spin-orbit coupling plays a pivotal role.
This state of matter is coined the quantum spin-Hall phase, and exhibits 
spin-filtered edge states carrying opposite currents for 
opposite spins, while possessing a bulk energy gap. Extensions of these findings
for finite Zeeman~\cite{sheng-2011} and magnetic~\cite{goldman-2012} fields in 
${\rm SU(2)}$ graphenelike lattices have indicated that quantum spin-Hall phases 
can survive the breaking of time-reversal symmetry.

However, in spite of a deeper understanding of topological properties of 
${\rm SU(2)}$ fermions in two-dimensional lattices found in 
condensed matter systems, very little is known about the topological properties 
of neutral ${\rm SU(N \ge 3)}$ fermions, such as $^{173}{\rm Yb}$, that have 
been loaded into optical 
lattices~\cite{takahashi-2012, takahashi-2018, fallani-2014, fallani-2016, 
bloch-2014a, bloch-2016}. 
Many experiments involving cold atoms have focused on studying 
topological properties of neutral ${\rm SU(2)}$ systems due to their direct
connections to their charged ${\rm SU(2)}$ cousins. A few experiments
have attempted to explore neutral ${\rm SU(2)}$ systems in fictitious magnetic 
and spin-orbit fields with the goal of studying the analogues of the quantum
charge-charge (charge-Hall) and spin-charge (spin-Hall) 
effects~\cite{bloch-2013, ketterle-2013}.

In this paper, we show that neutral ${\rm SU(3)}$ fermions 
are qualitatively different from their neutral or charged 
${\rm SU(2)}$ relatives. By labeling the internal states of the atoms as colors, 
we find that the topological insulating phases are characterized
by a set of three topological invariants: charge-charge, color-charge,
and color-color Chern numbers. This is in contrast with ${\rm SU(2)}$ systems,
where only charge-charge (charge-Hall) and spin-charge (spin-Hall) Chern numbers 
are necessary to classify topological insulators~\cite{sheng-2011}. 
We show examples of our classification for phase diagrams of 
neutral ${\rm SU(3)}$ fermions loaded into two-dimensional lattices and 
in the presence of fictitious magnetic, color-flip and color-orbit fields.

{\it Hamiltonian:} 
To describe topological phases of ${\rm SU(3)}$ fermions loaded into 
two-dimensional optical lattices, we start from the 
Hamiltonian 
\begin{equation}
\label{eqn:hamiltonian-second-quantization}
{\hat H} 
= 
-
\sum_{{\bf r}, {\boldsymbol \eta}_{\ell}}
t_{\ell}
\psi^\dagger ({\bf r})
e^{-i\boldsymbol \theta_{\ell}}
\psi ({\bf r} + {\boldsymbol \eta}_\ell)
-h_x 
\sum_{{\bf r}}
\psi^\dagger ({\bf r})
{\bf J}_x
\psi ({\bf r}),
\end{equation}
where $t_{\ell}$ are hopping energies along the $\ell = \{x, y\}$ direction,
and $h_x$ plays the role of
a color-flip field along the $x$ direction.
The phase operators $\boldsymbol \theta_{\ell}$ describe the 
effect of artificial color-orbit coupling 
${\boldsymbol \theta}_x = k_T \eta_x {\bf J}_z$,
with momentum transfer $k_T$
and artificial gauge field
${\boldsymbol \theta}_y = {\cal A}_y \eta_y {\bf 1}$, 
where ${\cal A}_y = e H_z x/\hbar c$ plays the role 
of the $y$ component of an artificial vector potential with dimension
of inverse length. Here, $H_z$ is identified as a synthetic 
magnetic field along the $z$-axis. 
The vector potential ${\cal A}_y$ may be generated by 
laser assisted tunneling~\cite{bloch-2013, ketterle-2013}, while the 
color-dependent momentum transfer $k_T$ and color-flip field $h_x$ may be 
created via counter-propagating Raman beams~\cite{spielman-2011} or via 
radio-frequency chips~\cite{spielman-2010}.
The vectors ${\boldsymbol \eta}_\ell = \eta_\ell {\boldsymbol {\hat \ell}}$,
with $\eta_{x} = \pm a_x$ and $\eta_{y} = \pm a_y$, indicate 
the position ${\bf r} + {\boldsymbol \eta}_\ell$ of nearest neighbors 
with respect ${\bf r} = (x, y)$, where fermion creation operators are  
defined by three-component vectors
$
\psi^\dagger ({\bf r}) 
= 
\left[
\psi^\dagger_{R} ({\bf r}),
\psi^\dagger_{G} ({\bf r}),
\psi^\dagger_{B} ({\bf r})
\right] 
$
with color $c = \{ R, G, B \}$ (Red, Green and Blue). 
The unit cell lengths are $a_x$ $(a_y)$ along the $x$ $(y)$ direction,
${\boldsymbol {\hat \ell}} = \{ {\hat {\bf x}}, {\hat {\bf y}} \}$ 
are the corresponding unit vectors, 
while the operators ${\bf J}_x$ and ${\bf J}_z$ are
pseudospin-1 matrices with states $\{ \uparrow, 0, \downarrow \}$ 
representing colors $\{ R, G, B \}$, respectively,
and ${\bf 1}$ is the identity matrix. 

Under the color-gauge transformation 
$
\psi ({\bf r}) = e^{i k_T x {\bf J}_z} {\widetilde \psi} ({\bf r}),
$
the Hamiltonian of Eq.~(\ref{eqn:hamiltonian-second-quantization}) 
becomes 
\begin{equation}
{\hat H} 
= 
-
\sum_{{\bf r}, {\boldsymbol \eta}_{\ell}}
t_{\ell}
{\widetilde \psi}^\dagger ({\bf r})
e^{-i{\widetilde {\boldsymbol \theta}}_{\ell}}
{\widetilde \psi} ({\bf r}) 
-h_x 
\sum_{{\bf r}}
{\widetilde \psi}^\dagger ({\bf r})
{\widetilde {\bf J}}
{\widetilde \psi} ({\bf r}),
\end{equation}
with
${\widetilde {\boldsymbol \theta}}_x  = 0$,
${\widetilde {\boldsymbol \theta}}_y  = {\boldsymbol \theta}_y$,
and 
${\widetilde {\bf J}} = {\bf J}_x \cos (k_T x) + {\bf J}_y \sin (k_T x)$.
When $h_x = 0$ the color-orbit coupling can be gauged
away (color-gauge symmetry), since the resulting Hamiltonian and 
its eigenvalues are independent of $k_T$. 

We transform the second quantization Hamiltonian of 
Eq.~(\ref{eqn:hamiltonian-second-quantization}) into the first quantization 
Hamiltonian matrix 
\begin{eqnarray} 
\label{eqn:hamiltonian-matrix}
{\hat {\bf H}} 
= 
\left(
\begin{array}{c c c }
\varepsilon_{R} ( \hat{\bf k}) & -h_x\sqrt{2}   &0  \\
-h_x\sqrt{2}   &  \varepsilon_{G} (\hat {\bf k}) & -h_x/\sqrt{2}\\\
 0              &  -h_x/\sqrt{2} &    \varepsilon_{B} (\hat {\bf k})
\end{array} 
\right) 
\end{eqnarray}
that describes a color generalization of 
the original Harper's Hamiltonian
for ${\rm SU(2)}$ fermions~\cite{harper-1955}. 
The matrix elements are 
\begin{equation}
\varepsilon_{R} (\hat{\bf k}) =
-2t_x \cos[({\hat k}_x - k_{T})a_x] 
-2t_y \cos[({\hat k}_y  - {\cal A}_y)a_y ]
\end{equation}
corresponding to the kinetic energy of the $R$ state, 
\begin{equation}
\varepsilon_{G} (\hat{\bf k}) =
-2t_x \cos[({\hat k}_x a_x)] 
-2t_y \cos[({\hat k}_y  -{\cal A}_y)a_y ]
\end{equation}
corresponding to the kinetic energy of the $G$ state,
and
\begin{equation}
\varepsilon_{B} (\hat{\bf k}) 
=
-2t_x \cos[({\hat k}_x + k_{T})a_x] 
-2t_y \cos[({\hat k}_y  -{\cal A}_y)a_y]
\end{equation}
corresponding to the kinetic energy of the $B$ state. 
Lastly, $h_x$ is a color-flip field along the $x$ direction, 
whose physical origin is a Rabi term that couples Red and Green,
as well as, Green and Blue internal states of the atom. The Hamiltonian matrix 
in Eq.~(\ref{eqn:hamiltonian-matrix}) acts on a three-color wavefunction 
$
{\boldsymbol \Psi}({\bf r}) 
= 
\left[ \Psi_{R} ({\bf r}), \Psi_{G}({\bf r}), \Psi_{B} ({\bf r})  \right]^T,
$
where $T$ indicates transposition.

Rewriting Eq.~(\ref{eqn:hamiltonian-matrix}) in terms of spin-1 matrices
${\bf J}_{\ell}$, with $\ell = \{x, y, z \}$, leads to
\begin{equation}
\label{eqn:hamiltonian-matrix-pseudo-spin-one}
{\hat {\bf H}} 
= 
\varepsilon_{G} ( \hat {\bf k} ) {\bf 1}
- h_x {\bf J}_x 
- h_z (\hat {\bf k}) {\bf J}_z 
+ b_z (\hat {\bf k} ){\bf J}_z^2 
\end{equation}
where 
$h_x$ plays the role of a Zeeman field along the $x$ axis in spin-space,
$
h_z ({\hat {\bf k}}) 
= 
\left[ 
\varepsilon_{B} ({\hat {\bf k}})
-
\varepsilon_{R} ({\hat {\bf k}})
\right]/2
$
represents momentum dependent Zeeman field along the
$z$ axis in spin-space, and 
$
b_z ({\hat {\bf k}})
= 
\left[
\varepsilon_{B} ({\hat {\bf k}})
+
\varepsilon_{R} ({\hat {\bf k}})
\right]/2
- 
\varepsilon_{G} ({\hat {\bf k}})
$
describes a momentum dependent quadratic Zeeman field
along the $z$ axis in spin-space. 
The explicit forms of the operators are 
$
h_z ({\hat {\bf k}})
=
2t_x \sin(k_T a_x) \sin({\hat k}_x a_x)
$
and
$
b_z ({\hat {\bf k}}) 
= 
4t_x \sin^{2}(k_T a_x/2) \cos({\hat k_x} a_x).
$
The term $b_z ({\hat {\bf k}}) {\bf J}_z^2$
describes a momentum dependent color-quadrupole (or pseudo-spin-quadrupole)
coupling, reflecting the entanglement of momentum and tensorial degrees
of freedrom~\cite{kurkcuoglu-2015, kurkcuoglu-2018a, kurkcuoglu-2018b}. 
The presence of the color fields $h_x$, $h_z (\hat {\bf k})$ and $b_z (\hat {\bf k})$ 
breaks ${\rm SU(3)}$ symmetry~\cite{footnote-SU3}, however 
the color-gauge transformation restores ${\rm SU(3)}$ symmetry 
when $h_x = 0$ for any value of $k_T$.
The term $b_z (\hat {\bf k}) {\bf J}_z^2$ is 
absent for ${\rm SU(2)}$ fermions in the presence of spin-orbit coupling, 
but, here, it plays a very important role in the determination 
and classification of topological insulating phases that emerge between  
degenerate ${\rm SU(3)}$ symmetric color insulators at $h_x = 0$ 
and fully polarized color insulators at $h_x \to \infty$. 

{\it Eigenspectrum:}
We choose first a cylindrical geometry with periodic boundary
conditions along the $y$ direction, and a finite number $M_x$ of sites
along the $x$ direction. In this case, 
$k_y$ is a good quantum number, while $k_x$ is not, leading to 
the color-dependent Harper's matrix 
\begin{eqnarray}
\label{eqn:hamiltonian-matrix-cylinder-geometry}
{\bf H} 
=
\left(
\begin{array}{c c c c c}
{\bf A}_{m-2}  & {\bf B}       &  {\bf 0}      &  {\bf 0}      & {\bf 0}       \\
{\bf B}^*      & {\bf A}_{m-1} &  {\bf B}      &  {\bf 0}      & {\bf 0}       \\
{\bf 0}        & {\bf B}^*     &  {\bf A}_m    & {\bf B}       & {\bf 0}       \\
{\bf 0}        & {\bf 0}       &  {\bf B}^*    & {\bf A}_{m+1} & {\bf B}       \\ 
{\bf 0}        & {\bf 0}       &  {\bf 0}      & {\bf B}^*     & {\bf A}_{m+2} \\
\end{array}
\right),
\end{eqnarray}
which has a tridiagonal block structure coupling neighboring sites 
$(m-1, m, m+1)$ along the $x$ direction, with $x = m a_x$, and discrete 
translational symmetry along the $y$ axis. The matrices
${\bf A}$, ${\bf B}$ and the null matrix ${\bf 0}$ 
consist of $3\times3$ blocks with entries labeled by internal 
color states $\{R, G, B\}$ or pseudo-spin-1 states 
$\{\uparrow, 0, \downarrow \}$. 
The size of the space labeled by the site index $m$ is $M_x$, thus 
the total dimension of the matrix ${\bf H}$ in 
Eq.~(\ref{eqn:hamiltonian-matrix-cylinder-geometry})
is $3M_x \times 3M_x$. The matrix indexed by position $x = m a_x$ is   
\begin{eqnarray}
{\bf A}_m 
=
\left(
\begin{array}{ccc}
{\bf A}_{m R} & -h_x/\sqrt{2}   & 0\\
-h_x/\sqrt{2}        & {\bf A}_{m G}    & -h_x/\sqrt{2}\\
0                   & -h_x/\sqrt{2}    & {\bf A}_{m B}
\end{array}
\right),
\nonumber
\end{eqnarray}
with
$
{\bf A}_{m R } 
= 
{\bf A}_{m G} 
= 
{\bf A}_{m B} 
= -2t_y \cos(k_y a_y - 2\pi m \alpha). 
$
Here, $\alpha = \Phi/\Phi_0$ is the ratio 
of the magnetic flux through a lattice plaquete $\Phi = H_z a_x a_y$ 
to the flux quantum $\Phi_0 = hc/e$. 
The matrix containing the color-orbit coupling is 
\begin{eqnarray}
{\bf B}
=
\left(
\begin{array}{ccc}
-t_x e^{-ik_{T}a_x} & 0    & 0  \\
0                & -t_x  & 0  \\
0                & 0     & -t_x e^{ik_{T}a_x}
\end{array}
\right),
\nonumber
\end{eqnarray}
where $k_T$ $(-k_T)$ corresponds to the momentum transfer along the 
$x$ direction for state $R$ $(B)$, while the momentum transfer for state
$G$ is zero.

\begin{figure}[tb]
\centering 
\epsfig{file=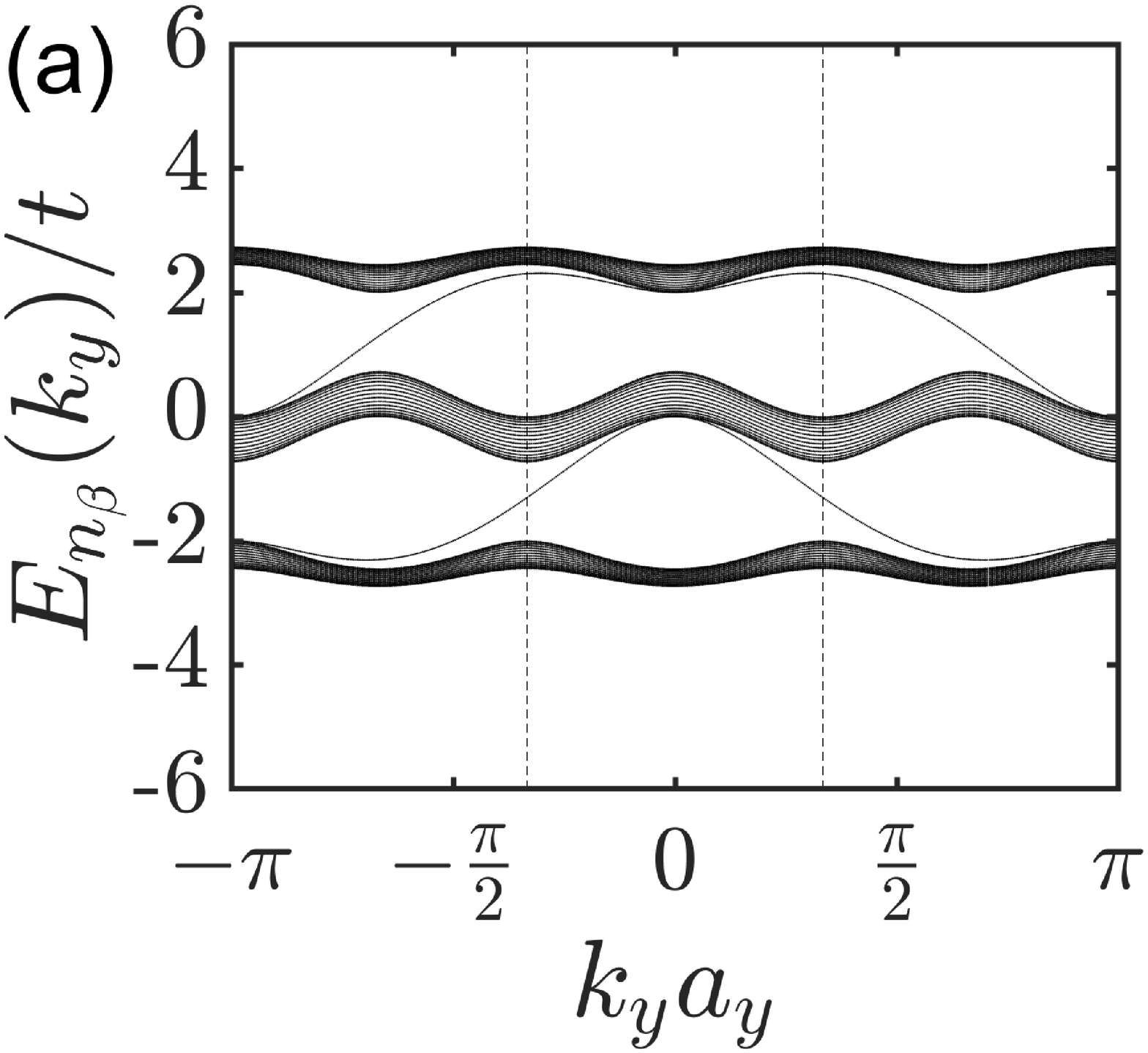,width=0.49 \linewidth}
\epsfig{file=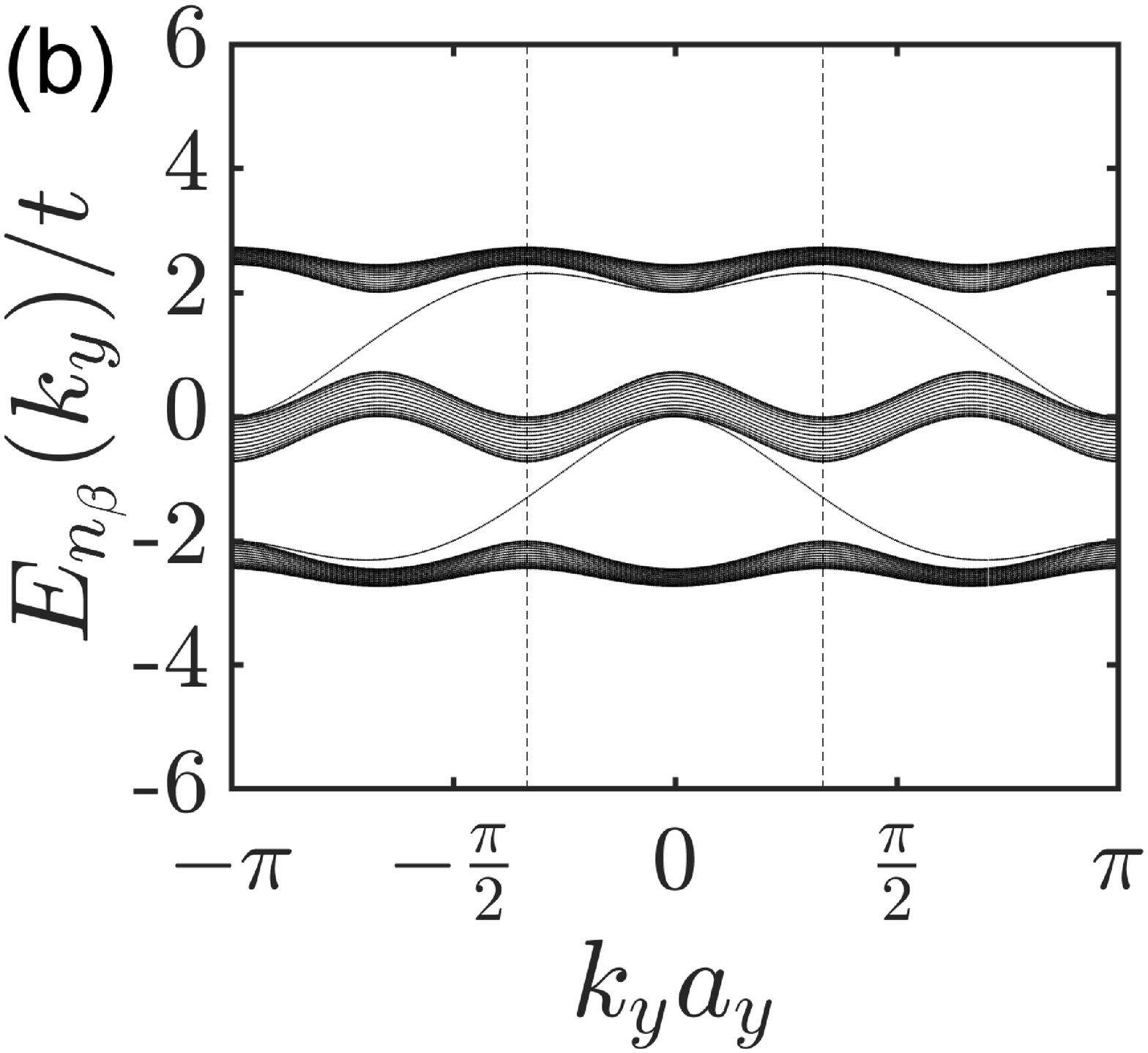,width=0.49 \linewidth}
\vskip 0.2cm
\epsfig{file=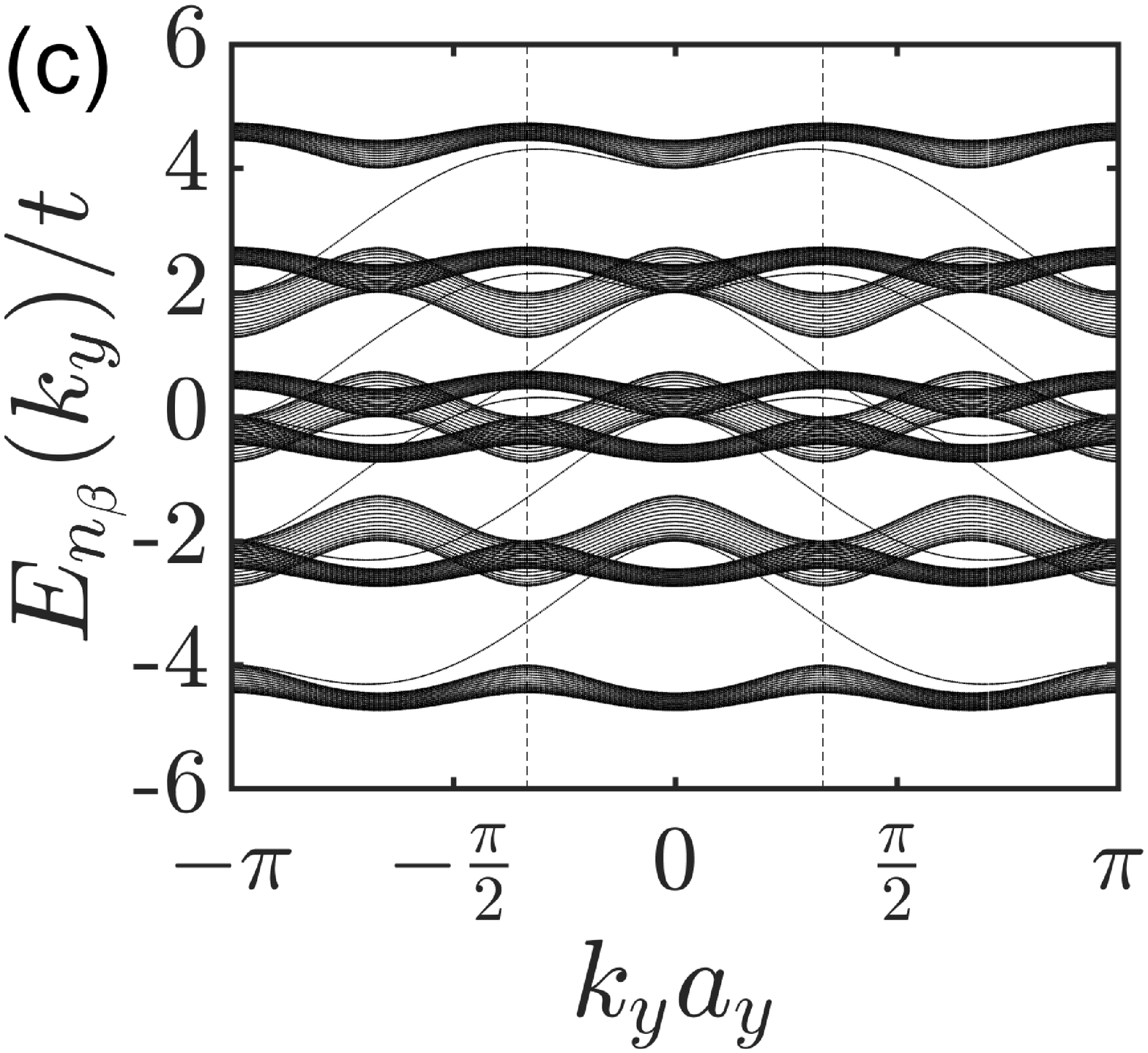,width=0.49 \linewidth}
\epsfig{file=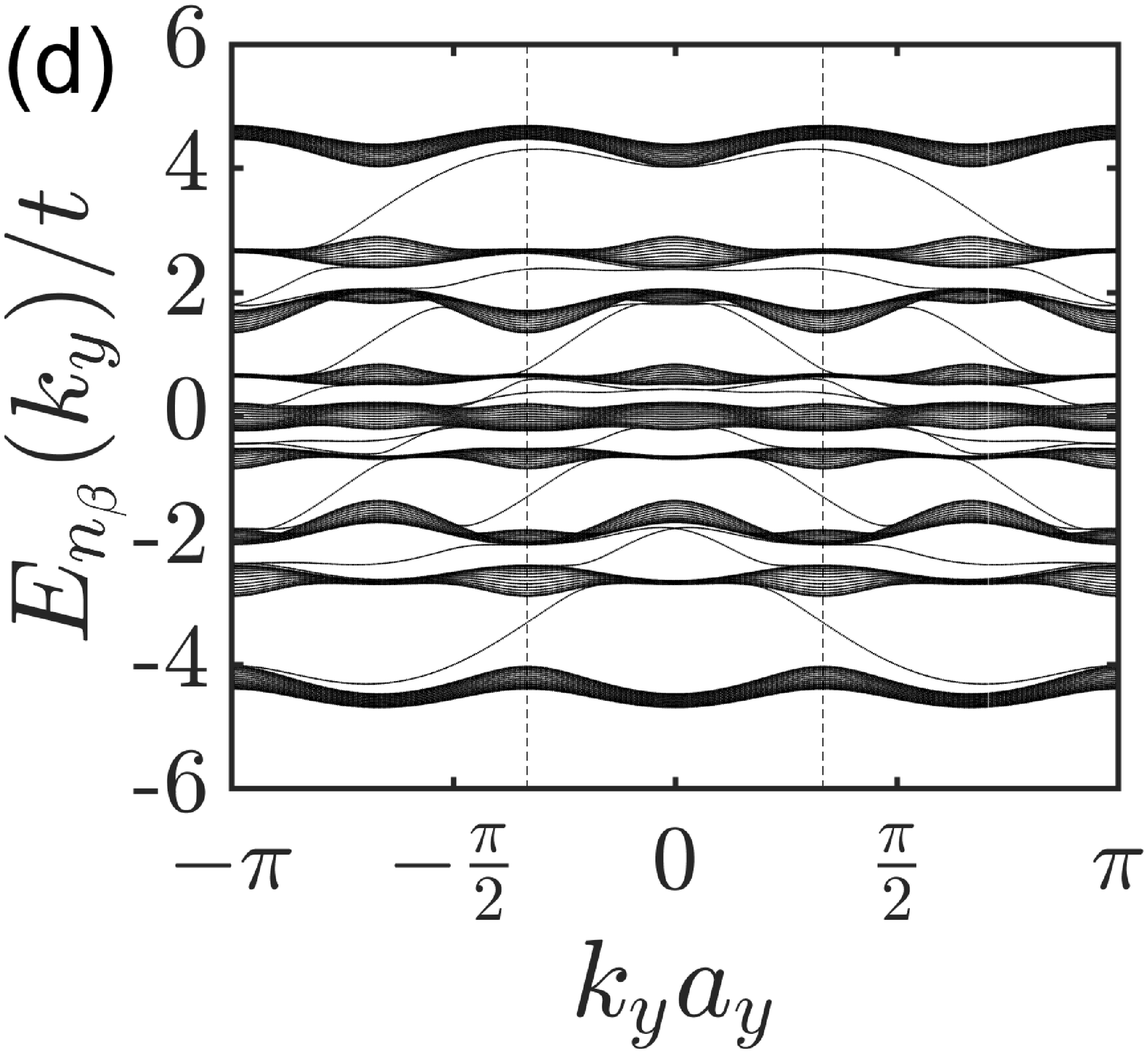,width=0.49 \linewidth}
\caption{ 
\label{fig:one}
Eigenvalues $E_{n_\beta} (k_y)$ of the color-dependent Harper's matrix
versus $k_y a_y$ for magnetic flux ratio $\alpha = 1/3$
and $t_x = t_y$. 
The parameters are: 
(a) $k_T a_x = 0$ and $h_x/t_y = 0$,  
(b) $k_T a_x = \pi/8$ and $h_x/t_y = 0$, 
(c) $k_T a_x = 0$ and $h_x/t_y = 2$, 
(d) $k_T a_x  = \pi/8$ and $h_x/t_y = 2$. 
The vertical dashed lines located at $k_y a_y = \pm \pi/3$ indicate the 
boundaries of the magnetic Brillouin zone. 
The bulk bands have periodicity $2\pi/3a_y$, and the 
edge bands have periodicity $2\pi/a_y$ along the $k_y$ direction.
}
\end{figure}

We consider $M_x = 50$ sites along the $x$ direction, with three 
states $\{R, G, B\}$ per site, 
but periodic boundary conditions along the $y$ direction. 
The eigenvalues $E_{n_\beta}( k_y )$ are labeled by a discrete band 
index $n_{\beta}$ and by momentum $k_y$, and are functions of the color-orbit coupling 
$k_T$, color-flip field $h_x$ and flux ratio $\alpha = \Phi/\Phi_0$.  
In Fig.~\ref{fig:one}, we show $E_{n_\beta} (k_y)$ for flux ratio $\alpha = 1/3$  
in four cases.
In Fig.~\ref{fig:one}(a) with $k_T a_x = 0$ and $h_x/t_y = 0$, 
there are three sets of degenerate bulk bands connected by color-degenerate 
edge states. 
In Fig.~\ref{fig:one}(b) with $k_T a_x = \pi/8$ and $h_x/t_y = 0$, the plots are 
identical to case (a) because of the color-gauge symmetry allows gauging away 
the color-orbit coupling. Notice in (a) and (b) that the bulk band gaps 
are connected by edge states at filling factors $\nu = \{1, 2\}$.
In Fig.~\ref{fig:one} with (c) $k_T a_x = 0$ and $h_x/t_y = 2$, there are nine sets 
of bulk bands with regions of overlap because color-degeneracies are 
only partially lifted by the color-flip field, and the bulk bands are also 
connected by color-dependent edge states. Bulk gaps are now present
at $\nu = \{1/3, 1, 2, 8/3\}$. 
In Fig.~\ref{fig:one}(d) with $k_T a_x = \pi/8$ and $h_x/t_y = 2$, there 
are nine sets of bulk bands connected by color-dependent edge states, 
but residual bulk band overlaps are lifted by color-orbit coupling, that is, 
because $k_T a_x \ne 0$.
Therefore, new insulating states arise at filling factors
$\nu = \{ 2/3, 4/3, 5/3, 7/3 \}$ due to the presence 
of $h_z (\hat {\bf k})$ and $b_z (\hat {\bf k})$ in 
Eq.~(\ref{eqn:hamiltonian-matrix-pseudo-spin-one}).

Each eigenvalue $E_{n_\beta} (k_y)$ has associated 
eigenstates $\vert k_y, n_\beta, \beta \rangle$, where $\beta$ is a mixed-color
index, reflecting the mixing of color components induced by $h_x$ and $k_T$.
The eigenstates can be written as a linear combination 
$
\vert k_y, n_{\beta}, \beta \rangle 
= 
\sum_{n c_x} u_{n_\beta n}^{\beta c_x} \vert k_y, n, c_x \rangle,
$
where $c_x$ represents the color basis with quantization axis along $h_x$, 
and $n$ is the band index in the absence of color-orbit coupling. 
The $\beta$ index has
three assigned values $\{\Uparrow, \emptyset, \Downarrow \}$ or
$\{C, M , Y \}$ (Cyan, Magenta, Yellow) to indicate the mixed-color nature
of the state. 

{\it Chern matrix:}
To characterize the topological nature of the insulating
phases, we impose the generalized boundary condition on the many-particle
wavefunction
$
\Psi ( {\bf r}_{1c}, ... , {\bf r}_{j c} + {\bf L}_{\ell}, ... ,
{\bf r}_{N_p c} )
=
e^{i\phi_{\ell c}}
\Psi ( {\bf r}_{1c}, ... , {\bf r}_{j c}, ... ,{\bf r}_{N_p c} ),
$
where ${\bf r}_{j c}$ is the position of the $i^{th}$ particle 
of color $c$, $N_p$ is the total number of particles,
${\bf L}_{\ell} = M_{\ell} a_{\ell} {\boldsymbol {\hat \ell}}$ 
is the length vector along $\ell = \{ x, y \}$ direction, 
and $\phi_{\ell}^c$ is the phase twist~\cite{niu-1985} along $\ell$ for color $c$.
Under the transformation
$
{\widetilde \Psi} ( {\bf r}_{1c}, ... , {\bf r}_{j c}, ... , {\bf r}_{N_p c} )
=
e^{-i \sum_{j, c}
\left [
\phi_{x c} \frac{x_{jc}}{M_x a_x}
+
\phi_{y c} \frac{y_{jc}}{M_y a_y}
\right]
}
\Psi ( {\bf r}_{1c}, ... , {\bf r}_{j c}, ... , {\bf r}_{N_p c} )
$
with $0 \le \phi_{{\ell} c} < 2\pi$, the wavefunction
$
{\widetilde \Psi} ( {\bf r}_{1c}, ... , {\bf r}_{j c}, ... , {\bf r}_{N_p c} )
$
is periodic in $\phi_{{\ell} c}$, and we can define 
the Chern matrix~\cite{sheng-2003} 
\begin{equation}
\label{eqn:chern-matrix}
{\cal C}_{cc^\prime}
=
\frac{i}{4\pi}
\iint
d\phi_{x c} d\phi_{y c^\prime}
{\cal F}_{xy} (\phi_{x c}, \phi_{y c^\prime}),
\end{equation}
where the purely imaginary {\it curvature} function 
\begin{equation}
{\cal F}_{xy} (\phi_{x c}, \phi_{y c^\prime})
=
\Bigg\langle \frac{\partial {\widetilde \Psi}}{\partial \phi_{x c}}
\Bigg\vert
\frac{\partial {\widetilde \Psi}}{\partial \phi_{y c^\prime}} 
\Bigg\rangle
-
\Bigg\langle \frac{\partial {\widetilde \Psi}}{\partial \phi_{y c^\prime}}
\Bigg\vert
\frac{\partial {\widetilde \Psi}}{\partial \phi_{x c}} 
\Bigg\rangle
\end{equation}
is integrated over the torus ${\cal T}^2_{c c^\prime}$, that is, over
the ranges of phase twists $0 \le \phi_{x c} < 2\pi$ 
and $0 \le \phi_{y c^\prime} < 2\pi$. The dimension of 
the Chern number matrix is $3 \times 3$, since there are three 
color states $c = \{ R, G, B \}$ or
three pseudo-spin 1 states $c = \{\uparrow, 0, \downarrow \}.$
In passing, we note that the SU(N) generalization for $N > 3$ leads to an 
$N \times N$ Chern matrix.

The expression given in Eq.~(\ref{eqn:chern-matrix}) is an integer 
just like in ${\rm SU(2)}$ systems~\cite{haldane-2006}.  
Three topological invariants can be obtained from the Chern matrix above. 
The first invariant is the charge-charge (charge-Hall) Chern number  
$
C_{\rm ch}^{\rm ch} 
= 
\sum_{c c^\prime}{\cal C}_{c c^\prime}.
$
The second is the color-charge (color-Hall) Chern number 
$
C_{\rm co}^{\rm ch} 
= 
\sum_{c c^\prime} m_c {\cal C}_{c c^\prime},
$
or 
charge-color Chern number
$
C_{\rm ch}^{\rm co} 
= 
\sum_{c c^\prime}  {\cal C}_{c c^\prime} 
m_{c^\prime},
$
since 
$
C_{\rm co}^{\rm ch} = 
C_{\rm ch}^{\rm co}. 
$
The third topological invariant is 
the color-color Chern number
$
C_{\rm co}^{\rm co} 
= 
\sum_{c c^\prime} m_c {\cal C}_{c c^\prime}m_{c^\prime},
$ 
where $m_c$ is the color quantum number with $m_R = +1$, $m_G = 0$, and 
$m_B = -1$, as identified from the pseudo-spin 1 representation 
$m_\uparrow = +1$, $m_0 = 0$, and $m_\downarrow = -1$. 

A simple way to connect these results to conventional ${\rm SU(2)}$ 
condensed matter physics of electrons and holes is to look at the
current density $J_x^{\lambda}$, where $\lambda$ refers to either 
charge or color, that is, $\lambda = \{ {\rm ch}, {\rm co} \}$ 
and the conductivity tensor ${\widetilde \sigma}_{xy}^{\lambda \tau}$ 
through the generalized relation 
$J_x^{\lambda} = {\widetilde \sigma}_{xy}^{\lambda \tau} E_y^{\tau}$,
where $E_y^{\tau}$ plays the role of a generalized {\it electric} field
with $\tau = \{ {\rm ch}, {\rm co} \}$.
To simplify our notation we drop the $xy$ labels, define the conductivity tensor
${\widetilde \sigma}_{xy}^{\lambda \tau} \equiv {\widetilde \sigma}_{\tau}^{\lambda}$,
and work finally with the conductance tensor 
$\sigma_{xy}^{\lambda \tau} \equiv \sigma_{\tau}^{\lambda}$.
If we were dealing with fermions with charge $e$ and conserved
pseudo-spin 1 projection along a global quantization axis, then 
the charge-charge (charge-Hall) conductance would be 
$
\sigma_{\rm ch}^{\rm ch} = (e^2/h) C_{\rm ch}^{\rm ch},
$
the color-charge (color-Hall) conductance would be 
$
\sigma_{\rm co}^{\rm ch} 
= 
(e^2/h) (\hbar/e) C_{\rm co}^{\rm ch} 
= 
(e/2\pi) C_{\rm co}^{\rm ch}
$ 
and the color-color conductance would be 
$
\sigma_{\rm co}^{\rm co}
= (e/2\pi) (\hbar/e) C_{\rm co}^{\rm co}
= (\hbar/2\pi) C_{\rm co}^{\rm co}.
$ 
However, our fermions are really neutral and their colors represent 
three internal states of the atoms, thus one can only hope to 
probe the charge-charge (charge-Hall) and color-charge (color-Hall) 
and color-color Chern numbers in analogy 
with measurement proposals~\cite{satija-2011, cooper-2012, goldman-2013} 
or actual measurements~\cite{esslinger-2014, bloch-2015} 
of Chern numbers for atomic systems with one and two internal states.

In the ${\rm SU(2)}$ case, the topological invariant equivalent to 
$
C_{\rm co}^{\rm co}
$ 
is the spin-spin Chern number 
$
C_{\rm sp}^{\rm sp} 
= \sum_{\sigma} m_{\sigma}^2 C_{\sigma \sigma}
$
which has the same value as charge-charge (charge-Hall) Chern number 
$
C_{\rm ch}^{\rm ch} 
= 
\sum_{\sigma} C_{\sigma \sigma},
$ 
since $m_{\sigma}^2 = 1$.
Since $C_{\rm sp}^{\rm sp}$ does not add any additional information about
the topological nature of insulating phases for ${\rm SU(2)}$ systems, it 
is sufficient to stop the topological classification at the spin-charge 
(spin-Hall) level, such as the ${\rm Z_2}$ classification used in the case of 
quantum spin-Hall phases of graphene-like structures~\cite{kane-2005, sheng-2011}.
However, in the ${\rm SU(3)}$ case, $C_{\rm co}^{\rm co}$
provides new topological information and can be used to refine 
the topological classification of the non-trivial insulating phases.
We note that for ${\rm SU(N)}$ fermions with
$N > 3$ flavors, the generalized flavor-flavor Chern number 
$C_{\rm fl}^{\rm fl}$ will also provide additional topological information 
about the insulating states. 

\begin{figure} [tb]
\centering 
\epsfig{file=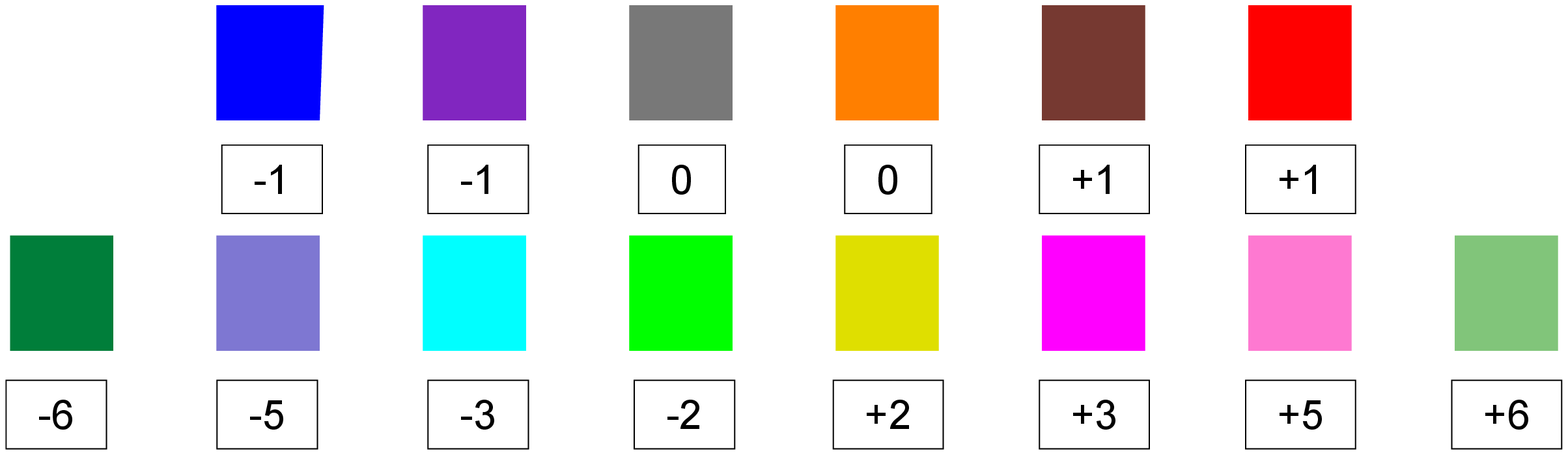, width = 0.98\linewidth}
\vskip 0.2cm
\epsfig{file=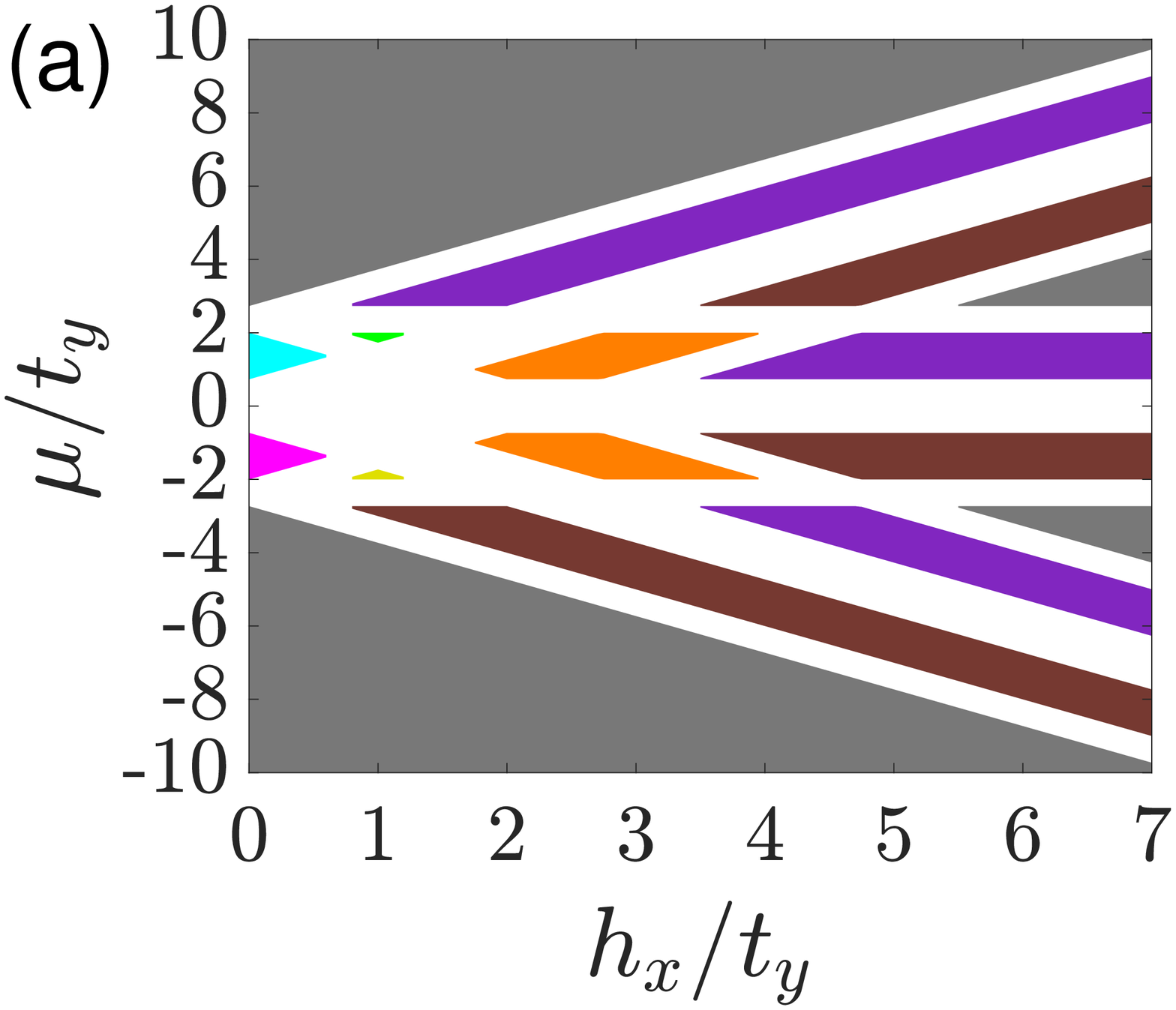,width= 0.49 \linewidth}
\epsfig{file=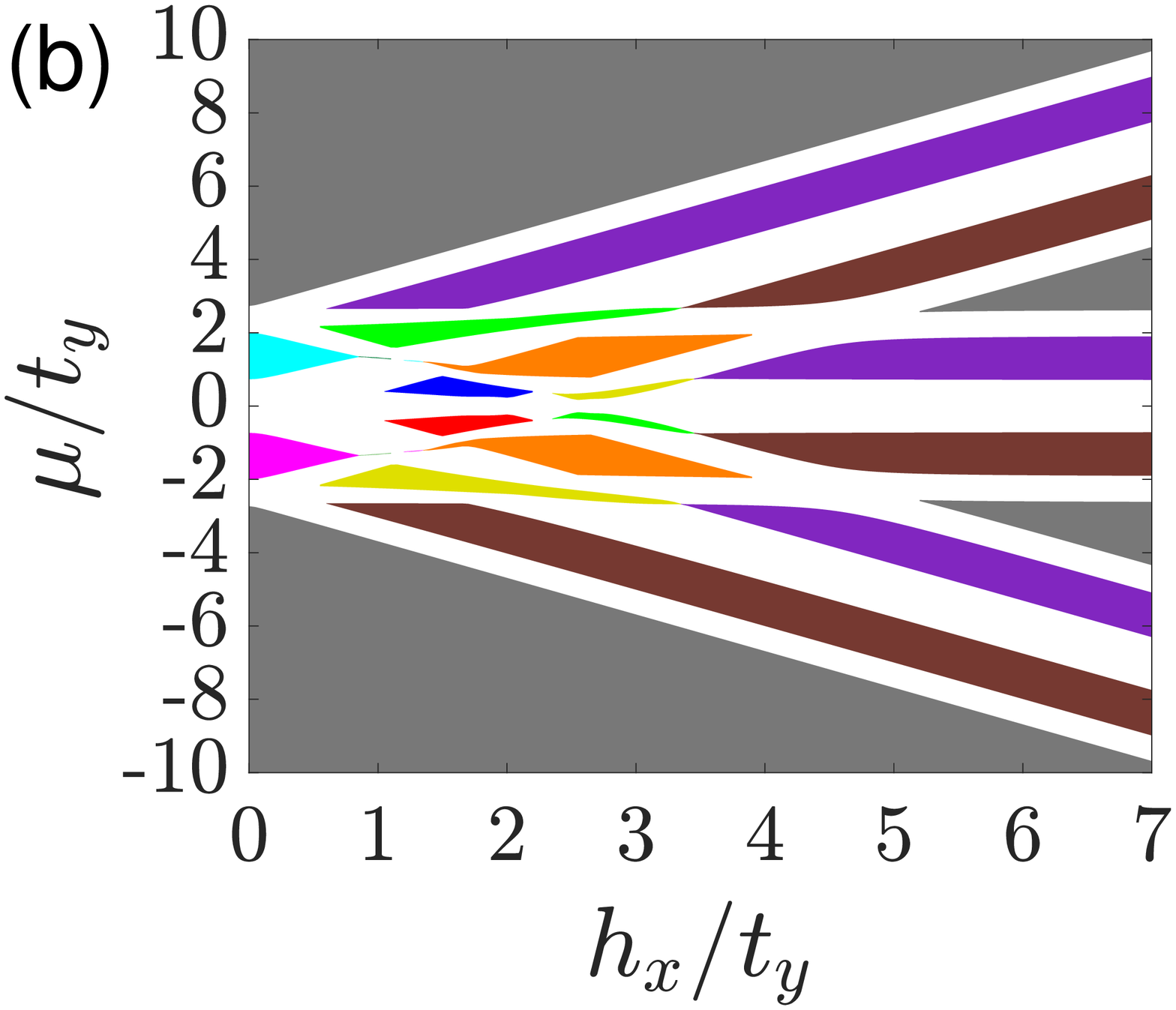,width= 0.49\linewidth}
\caption{ 
\label{fig:two}
Phase diagrams of chemical potential $\mu/t_y$ versus
Zeeman field $h_x/t_y$ for $t_x = t_y$ and
color-orbit coupling parameters: 
(a) $k_T a_x = 0$, 
(b) $k_T a_x = \pi/8$. 
The white (non-white) regions correspond to conducting 
(insulating) phases. 
The values of the charge-charge (charge-Hall) Chern numbers 
$C_{\rm ch}^{\rm ch}$ for each insulating region 
are indicated in the legend. 
}
\end{figure}

{\it Phase Diagrams:} 
In Fig.~\ref{fig:two}, we show the phase diagram of chemical potential
$\mu/t_y$ versus Zeeman field $h_x/t_y$ for fixed value of the 
magnetic flux ratio $\alpha = 1/3$, hoppings $t_x = t_y$ and two values of 
the color-orbit parameter:
(a) $k_T a_x = 0$, and (b) $k_T a_x = \pi/8$.
The white regions indicate conducting phases, while the regions 
with other colors correspond to insulating phases. The color palette 
in Fig.~\ref{fig:two} indicates the charge-charge (charge-Hall) 
Chern numbers $C_{\rm ch}^{\rm ch}$ associated with 
the corresponding colored regions.  However,  
$C_{\rm ch}^{\rm ch}$ is not sufficient to classify the topological insulating 
phases of ${\rm SU(3)}$ fermions for arbitrary $k_T a_x$ and $h_x/t_y$, 
as we also need the color-charge (color-Hall) $C_{\rm co}^{\rm ch}$ and
color-color $C_{\rm co}^{\rm co}$ Chern numbers.

\begin{figure} [tb]
\centering 
\vskip 0.2cm
\epsfig{file=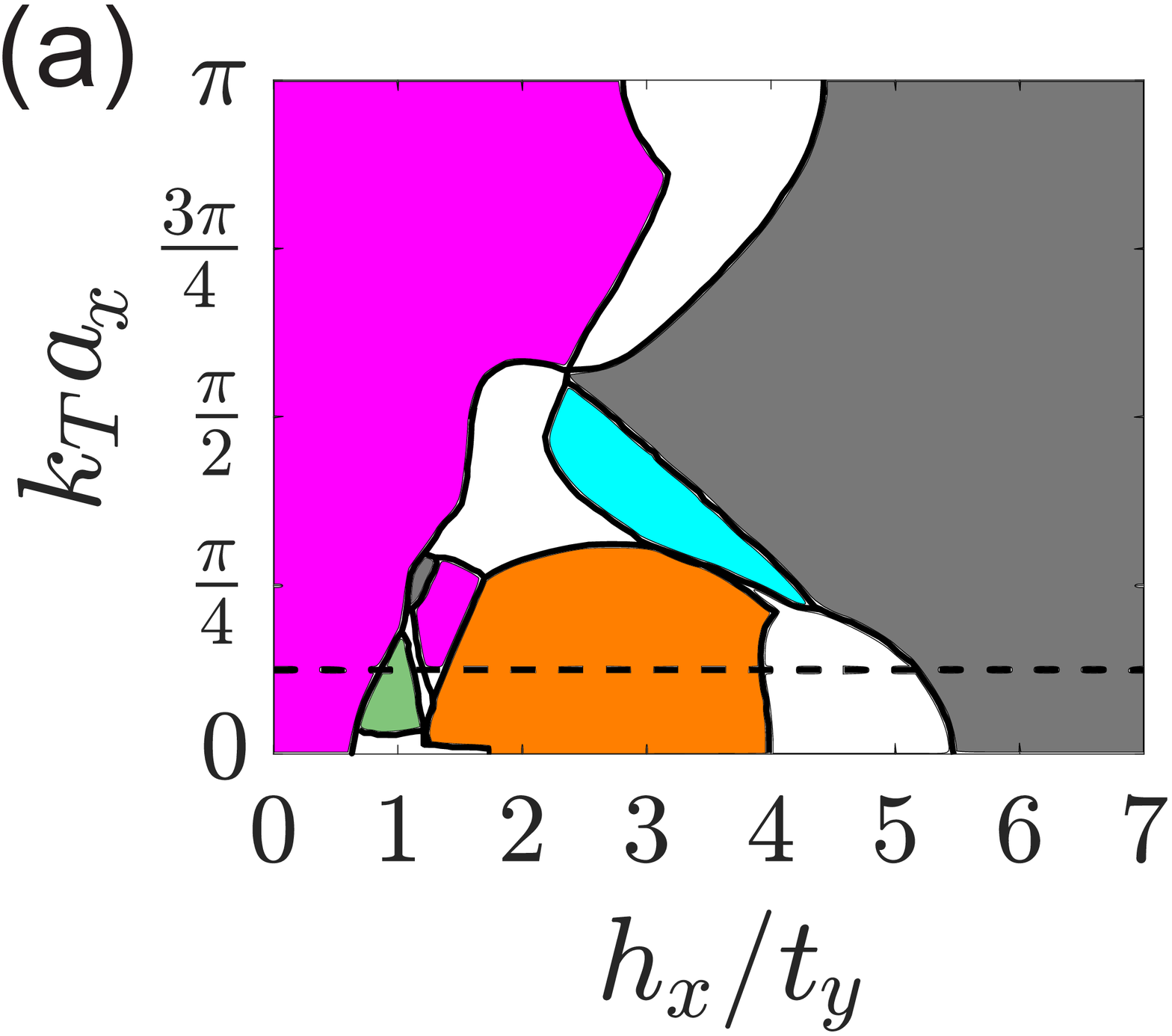,width=0.49 \linewidth}
\epsfig{file=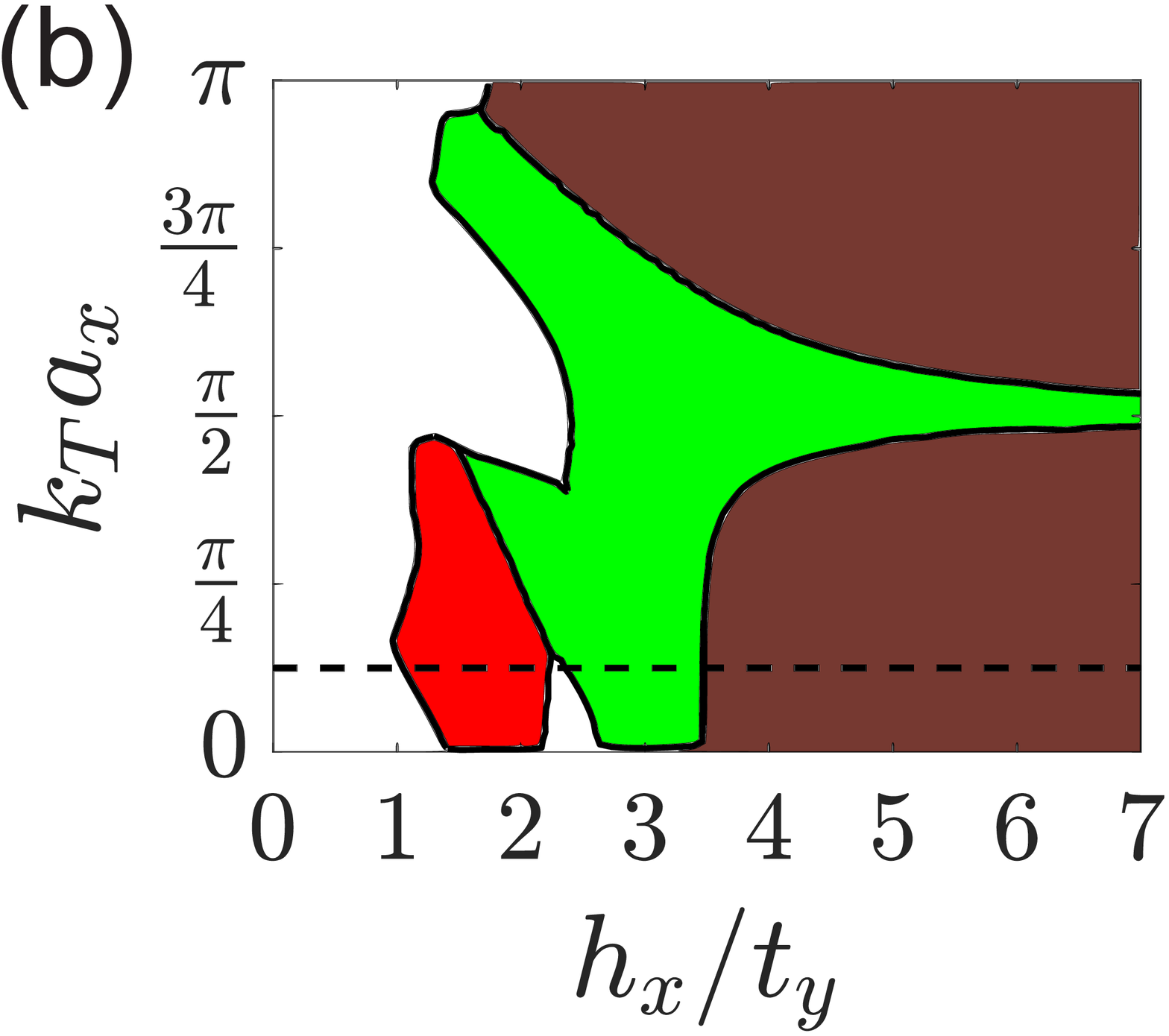,width=0.49 \linewidth}
\vskip 0.2cm
\epsfig{file=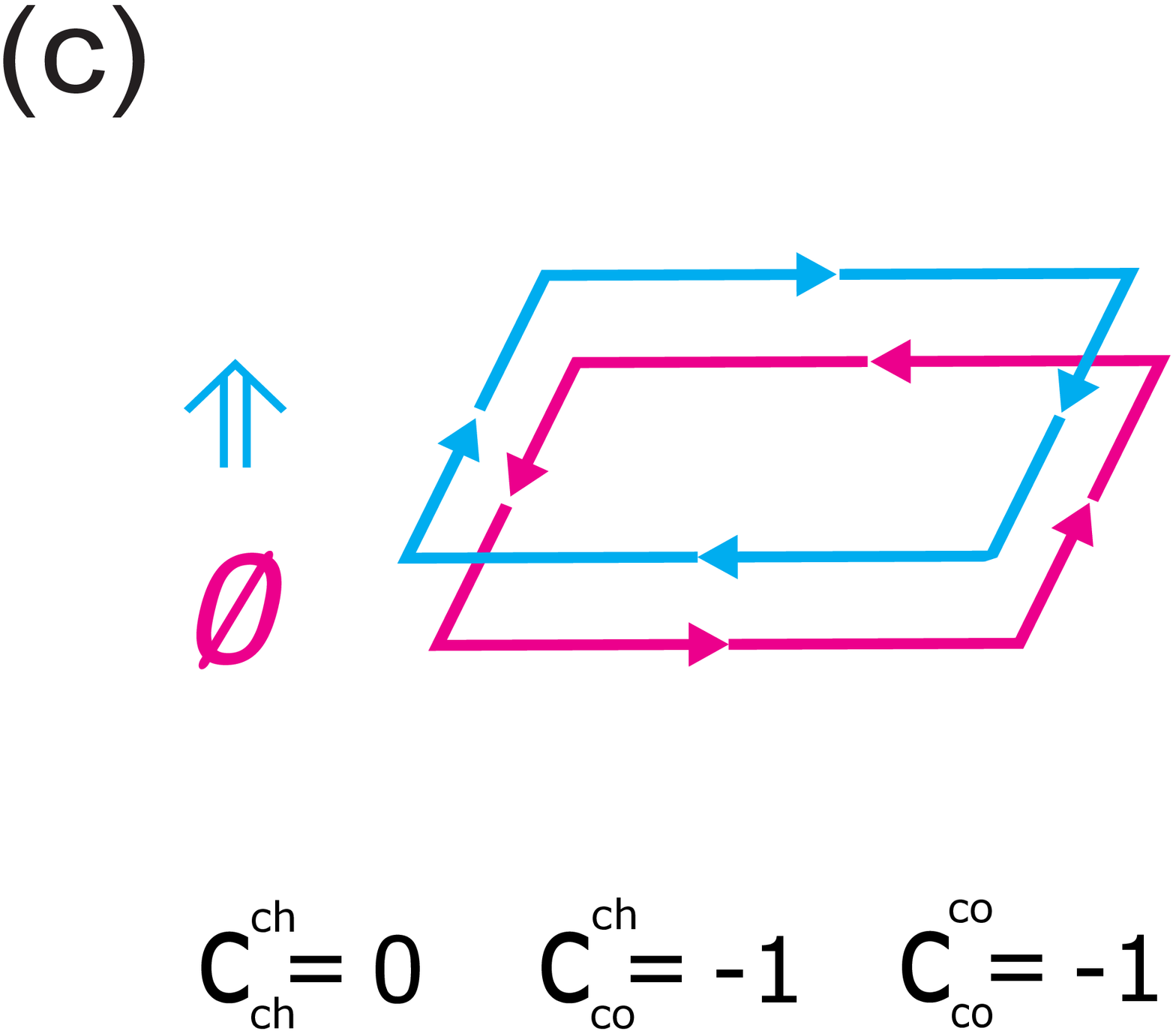,width=0.49 \linewidth}
\epsfig{file=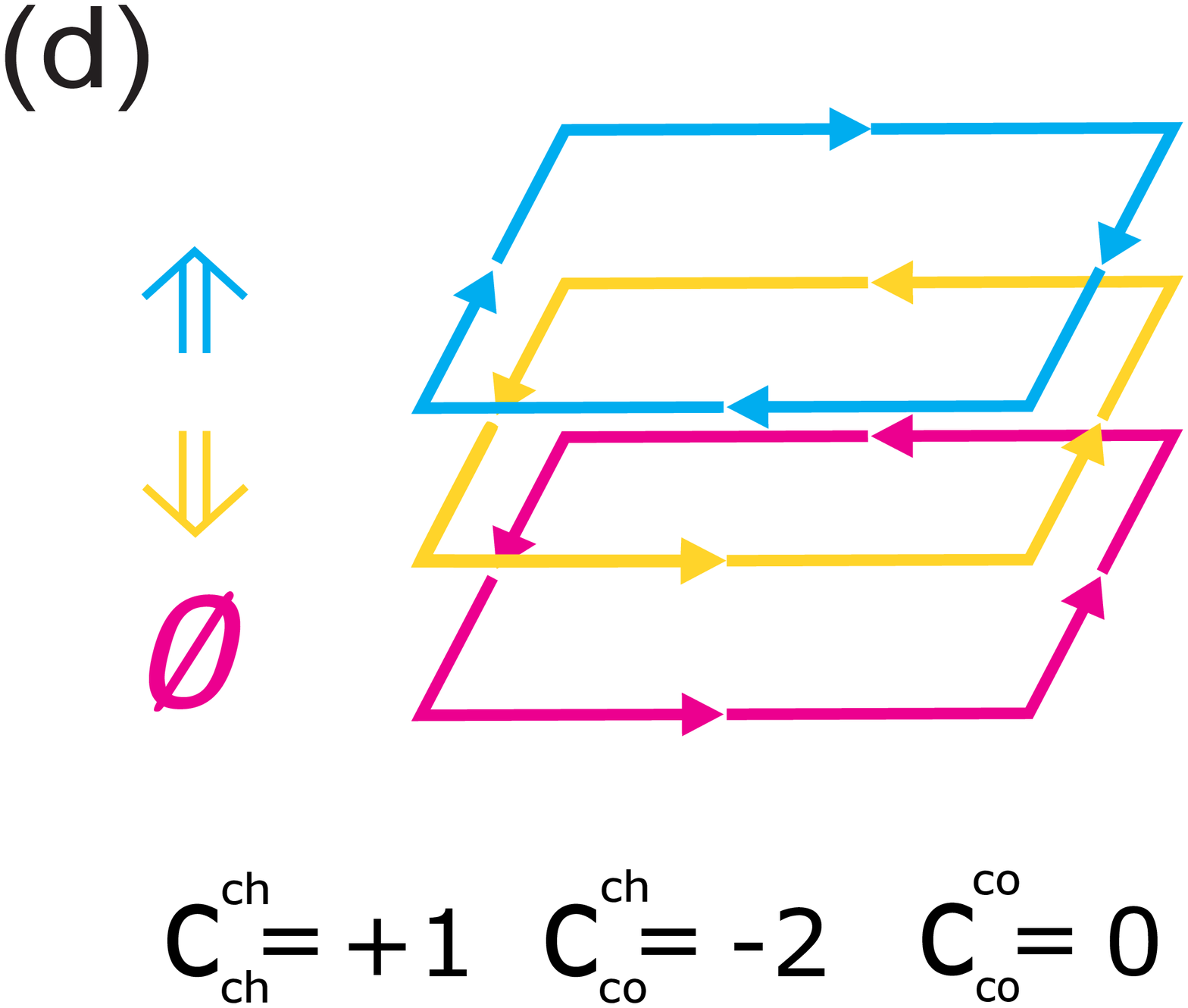,width=0.49 \linewidth}
\caption{
\label{fig:three}
(Color Online) 
Phase diagram of color-orbit coupling $k_T a_x$ versus color-flip field
$h_x/t_y$ for filling factors $\nu = 1$ in (a) and $\nu = 4/3$ in (b).
Dashed lines are shown for $k_T a_x = \pi/8$. 
The color coding is the same used in Fig.~\ref{fig:two}. 
In (c), edge states are shown for the orange region in (a). 
In (d), edge states are shown for the red region in (b).
Corresponding Chern numbers are listed.
}
\end{figure}

In Figs.~\ref{fig:two}(a) and~\ref{fig:two}(b) , we see our classification at work.
The gray regions, occuring at filling factors $\nu = 0, 1, 2,3$, 
are topologically trivial with either non-chiral or no edge states at all.
They have charge-charge (charge-Hall) Chern number $C_{\rm ch}^{\rm ch} = 0$ and 
also zero color-charge (color-Hall) and color-color Chern numbers 
$C_{\rm co}^{\rm ch} = C_{\rm co}^{\rm co} = 0$.
The magenta region at $\nu = 1$ has Chern numbers
$C_{\rm ch}^{\rm ch} = +3$, $C_{\rm co}^{\rm ch} = 0$ and
$C_{\rm co}^{\rm co} = +2$, while
the cyan region at $\nu = 2$ has Chern numbers
$C_{\rm ch}^{\rm ch} = -3$, $C_{\rm co}^{\rm ch} = 0$ and
$C_{\rm co}^{\rm co} = -2$. These regions, occuring at low values 
of $h_x/t_y$, are the analogue of the traditional quantum Hall phases 
of ${\rm SU(2)}$ fermions.
However, the yellow region at $\nu = 2/3$ with
$C_{\rm ch}^{\rm ch} = +2$, $C_{\rm co}^{\rm ch} = +1$ and
$C_{\rm co}^{\rm co} = +1$, and 
green region at $\nu = 7/3$ with
$C_{\rm ch}^{\rm ch} = -2$, $C_{\rm co}^{\rm ch} = +1$ and
$C_{\rm co}^{\rm co} = -1$, have no counterparts for 
${\rm SU(2)}$ fermions. 

The orange regions, occuring at filling factors 
$\nu = 1, 2$, have 
$C_{\rm ch}^{\rm ch} = 0$, but, unlike the gray regions, they are topologically 
non-trivial. Each orange region has two chiral edge states and 
non-zero 
$C_{\rm co}^{\rm ch} = -1$, but they are
distinguished further by their 
color-color Chern number $C_{\rm co}^{\rm co}$. 
The orange region with $\mu < 0$ and $\nu = 1$ has 
$C_{\rm co}^{\rm co} = -1$, 
however, the orange region 
with $\mu > 0$ and $\nu = 2$ has $C_{\rm co}^{\rm co} = +1$.
These are quantum color-Hall states, that is, the ${\rm SU(3)}$ analogues 
of the time-reversal broken quantum spin-Hall states that occur 
in ${\rm SU(2)}$ fermions.

The phases at high values of $h_x/t_y$, represented by the brown and purple regions,
describe color-polarized insulating states.
The brown region at $\nu = 1/3$ has 
$C_{\rm ch}^{\rm ch} = +1$, $C_{\rm co}^{\rm ch} = +1$ and
$C_{\rm co}^{\rm co} = +1$; 
the brown region at $\nu = 4/3$ has
$C_{\rm ch}^{\rm ch} = +1$, $C_{\rm co}^{\rm ch} = 0$ and
$C_{\rm co}^{\rm co} = 0$; and
the brown region at $\nu = 7/3$ has
$C_{\rm ch}^{\rm ch} = +1$, $C_{\rm co}^{\rm ch} = -1$ and
$C_{\rm co}^{\rm co} = +1$.
The purple region at $\nu = 2/3$ has
$C_{\rm ch}^{\rm ch} = -1$, $C_{\rm co}^{\rm ch} = -1$ and
$C_{\rm co}^{\rm co} = -1$; 
the purple region at $\nu = 5/3$ has
$C_{\rm ch}^{\rm ch} = -1$, $C_{\rm co}^{\rm ch} = 0$ and
$C_{\rm co}^{\rm co} = 0$; and
the purple region at $\nu = /3$ has
$C_{\rm ch}^{\rm ch} = -1$, $C_{\rm co}^{\rm ch} = +1$ and
$C_{\rm co}^{\rm co} = -1$. 
In Fig.~\ref{fig:two}(b), there are additional insulating phases induced 
by color-orbit coupling,
such as the red region at $\nu = 4/3$ with
$C_{\rm ch}^{\rm ch} = +1$, $C_{\rm co}^{\rm ch} = -2$ and
$C_{\rm co}^{\rm co} = 0$;
the blue region at $\nu = 5/3$ has
$C_{\rm ch}^{\rm ch} = -1$, $C_{\rm co}^{\rm ch} = -2$ and
$C_{\rm co}^{\rm co} = 0$;
the green region at $\nu = 4/3$ has
$C_{\rm ch}^{\rm ch} = -2$, $C_{\rm co}^{\rm ch} = 0$ and
$C_{\rm co}^{\rm co} = 0$;
and the yellow region at $\nu = 5/3$ has
$C_{\rm ch}^{\rm ch} = +2$, $C_{\rm co}^{\rm ch} = 0$ and
$C_{\rm co}^{\rm co} = 0$.

We plot phase diagrams of color-orbit coupling 
parameter $k_x a_x$ versus color-flip field $h_x/t_y$ for $\nu = 1$ 
in Fig.~\ref{fig:three}(a) and for $\nu = 4/3$ in in Fig.~\ref{fig:three}(b). 
Two phases not yet discussed arise in Fig.~\ref{fig:three}(a), 
a light green region with 
$C_{\rm ch}^{\rm ch} = +6$, $C_{\rm co}^{\rm ch} = -1$ and
$C_{\rm co}^{\rm co} = +3$, and a cyan region with 
$C_{\rm ch}^{\rm ch} = -3$, $C_{\rm co}^{\rm ch} = -1$ and
$C_{\rm co}^{\rm co} = -1$, while no new phases arise in 
Fig.~\ref{fig:three}(b). In order to relate ${\rm SU(3)}$ fermions
to their ${\rm SU(2)}$ cousins, we show schematically edge
states in Fig.~\ref{fig:three}(c) linked to the orange region 
in Fig.~\ref{fig:three}(a), and edge states in Fig.~\ref{fig:three}(d)
linked to the red region Fig.~\ref{fig:three}(b).

{\it Conclusions:}
For ${\rm SU(3)}$ fermions in optical lattices,
we showed that the classification of topological color insulators
requires three topological invariants: the charge-charge, the color-charge and 
the color-color Chern numbers. We analyzed ${\rm SU(3)}$
fermions in the presence of artificial magnetic, color-flip and color-orbit fields,
and indicated that our classification transcends that of ${\rm SU(2)}$
fermions, where only charge-charge (charge-Hall) and spin-charge (spin-Hall) 
Chern numbers are necessary to characterize topological insulating phases. 
Our findings open an avenue for the exploration of topological insulators of 
${\rm SU (N \ge 3)}$ fermions with and without interactions, and also suggest that
such phases may be found in lattice quantum chromodynamics models.   

One of us (C.A.R.S.d.M.) would like to thank the support of 
the Galileo Galilei Institute for Theoretical Physics via a Simons
Fellowship, and of the International Institute of Physics via the 
Visitor's Program.


\begin{thebibliography}{2}
\bibitem{footnote-SU3}
We note that the Hamiltonian in Eq.~(\ref{eqn:hamiltonian-matrix-pseudo-spin-one}) 
in general does not commute with the Gell-Mann matrices ${\boldsymbol \lambda}_j$, 
which are the eight generators of SU(3). To see this explicitly, it is
sufficient to recall that the angular momentum matrices ${\bf J}_\ell$
can be written in terms of ${\boldsymbol \lambda}_ j$ as 
%
$
{\bf J}_x 
= 
\left(
{\boldsymbol \lambda}_1 + {\boldsymbol \lambda}_6
\right)/2;
$
%
%
$
{\bf J}_y 
= 
\left( 
{\boldsymbol \lambda}_2 + {\boldsymbol \lambda}_7
\right)/2;
$
%
and
%
$
{\bf J}_z 
= 
\left(
{\boldsymbol \lambda}_3 + \sqrt{3}{\boldsymbol \lambda}_8
\right)/2
$
%
and to show that the commutator 
%
$
[ 
{\hat {\bf H} }, {\boldsymbol \lambda}_j 
]
\ne 0.
$
\end{thebibliography}
\end{document}